\documentclass{nature}

\usepackage{amsmath}    
\usepackage{color}
\usepackage[10pt]{moresize}
\usepackage{lineno}
\usepackage{graphicx}
\usepackage{caption}
\DeclareCaptionLabelFormat{adja-page}{\hrulefill\\#1 #2 \emph{(previous page)}}
\usepackage{subfig}
\usepackage{dsfont}
\usepackage{amsfonts}
\usepackage{amssymb}
\usepackage{amscd}
\usepackage{enumerate}
\usepackage{epsfig}
\usepackage{epstopdf}
\usepackage{dcolumn}
\usepackage{bm}
\usepackage{amsthm}
\usepackage{amsfonts}
\usepackage{color} 
\usepackage[usenames,dvipsnames]{xcolor}
\usepackage{amsmath}
\usepackage{enumerate}
\usepackage{bbm}
\usepackage[colorlinks=true,citecolor=blue,urlcolor=black]{hyperref}

\begin{document}

\title{Towards satellite-based quantum-secure time transfer}

\author{Hui Dai\textsuperscript{1,2,*}, Qi Shen\textsuperscript{1,2,*}, Chao-Ze Wang\textsuperscript{1,2}, Shuang-Lin Li\textsuperscript{1,2}, Wei-Yue Liu\textsuperscript{1,2}, Wen-Qi Cai\textsuperscript{1,2}, Sheng-Kai Liao\textsuperscript{1,2}, Ji-Gang Ren\textsuperscript{1,2}, Juan Yin\textsuperscript{1,2}, Yu-Ao Chen\textsuperscript{1,2}, Qiang Zhang\textsuperscript{1,2}, Feihu Xu\textsuperscript{1,2,$\star$}, Cheng-Zhi Peng\textsuperscript{1,2,$\star$}, Jian-Wei Pan\textsuperscript{1,2,$\star$}}

\maketitle

\begin{affiliations}
    \item Hefei National Laboratory for Physical Sciences at Microscale and Department of Modern Physics, University of Science and Technology of China, Hefei 230026, China
    \item Shanghai Branch, CAS Center for Excellence and Synergetic Innovation Center in Quantum Information and Quantum Physics, University of Science and Technology of China, Shanghai 201315, China
    \\
    $\textsuperscript{*}$These authors contributed equally to this work. \\
    $\textsuperscript{$\star$}$Corresponding author: feihuxu@ustc.edu.cn; pcz@ustc.edu.cn; pan@ustc.edu.cn.
\end{affiliations}
\baselineskip24pt

\begin{abstract}
High-precision time synchronization for remote clocks plays an important role in fundamental science\cite{riehle2017optical,kolkowitz2016gravitational,marra2018ultrastable} and real-life applications\cite{corbett2013spanner, de2010synchronized}. However, the current time synchronization techniques\cite{mills1991internet,lewandowski1999gps} have been shown to be vulnerable to sophisticated adversaries\cite{humphreys2008assessing}. There is a compelling need for fundamentally new methods to distribute high-precision time information securely. Here we propose a satellite-based quantum-secure time transfer (QSTT) scheme based on two-way quantum key distribution (QKD) in free-space, and experimentally verify the key technologies of the scheme via the Micius quantum satellite. In QSTT, a quantum signal (e.g., single photon) is used as the carrier for both the time transfer and the secret-key generation, offering quantum-enhanced security for transferring time signal and time information. We perform a satellite-to-ground time synchronization using single-photon-level signals and achieve a quantum bit error rate of less than 1\%, a time data rate of 9 kHz and a time-transfer precision of 30 ps. These results offer possibilities towards an enhanced infrastructure of time-transfer network, whose security stems from quantum physics.
\end{abstract}

High-precision time-frequency transfer plays a critical role in current infrastructures surrounding its real-life applications\cite{corbett2013spanner, de2010synchronized}. The practical security for time-frequency transfer has become an increasingly important issue in applications\cite{kerns2014unmanned, bhatti2017hostile}. The widely employed time synchronization protocol today, i.e., global navigation satellite system (GNSS)\cite{lewandowski1999gps,allan1980accurate}, faces security threats such as GPS spoofing\cite{warner2003gps,humphreys2008assessing}. Beside microwave signals, although important progress has been made for optical time-frequency transfer to achieve high precision and stability\cite{smith2006two,samain2008time,giorgetta2013optical,predehl2012920, sinclair2018comparing}, the security issue remains unaddressed.

Existing theoretical studies have analyzed the security of time transfer based on possible man-in-the-middle attacks\cite{treytl2007traps,ullmann2009delay,tippenhauer2011requirements,jafarnia2012gps}. One-way time transfer was proven to be vulnerable to the time-delay attack\cite{tippenhauer2011requirements}, where the adversary randomly increases/decreases the propagation time of the time signal. In contrast, two-way time transfer was known as the more secure method, since the time-delay attack can be countered by verifying the round-trip time. In general, security in the latter is governed by three necessary conditions\cite{narula2018requirements} (see Table \ref{Tab:secure conditions}): (i) The two stations must transmit authenticated time signals and time data; (ii) propagation time must be irreducible to within a security alert limit; and (iii) the round-trip time must be measurable to one of the two stations to within the alert limit. Condition (i) is to counter the intercept-resend attack for time signals and data, while condition (ii) and (iii) are primarily used to defend against the time-delay attack (see Methods).

Quantum technology brings new perspectives for time transfer, such as quantum clock synchronization protocols\cite{jozsa2000quantum,ilo2018remote}, quantum-enhanced synchronization\cite{giovannetti2001quantum}, synchronization with entangled pairs\cite{valencia2004distant,marcikic2006free,ho2009clock,lee2019asymmetric}, and quantum networks of clocks with high accuracy\cite{komar2014quantum}.

Motivated by the information-theoretic security of quantum communication\cite{lo1999unconditional,shor2000simple,lo2014secure}, we proposed a satellite-based quantum-secure time transfer (QSTT) protocol, which can fully meet the three security conditions (see Table \ref{Tab:secure conditions}). Based on the Micius quantum satellite\cite{liao2017satellite,yin2017satellite,ren2017ground}, we experimentally verify the key technologies of satellite-based QSTT, including the time transfer using single-photon-level signals from satellite to ground, the generation of secure keys via QKD, the secure transmission of quantum-encrypted data and so forth. We demonstrated a quantum time-transfer precision of $\sim$30 ps by using single photons as timing signals for optical time transfer from satellite to ground. This precision is comparable to the state-of-the-art technique of T2L2 on Jason-2 that used strong laser pules\cite{samain2008time}, while the achieved timing data rate of $\sim$9 kHz is higher than T2L2 of 10 Hz. A summary of different types of time-transfer schemes is shown in the Supplementary Table 2.

Fig. \ref{twoway_QKD} shows the schematic diagram of QSTT. Assumed to have master clock $\mathbf{A}$, satellite Alice initiates the two-way time transfer with the ground station, Bob, who has the slave clock $\mathbf{B}$. The scheme is implemented through four basic steps.

\emph{Step 1.} Alice and Bob mutually transmit single photons over free space for both two-way QKD and two-way transmission of timing signals. We use the polarization-encoding BB84 protocol for QKD\cite{lo2014secure}. Alice prepares a single photon in the randomly polarized BB84 state and at time of $t^{A}_{S}$ by her local clock $\mathbf{A}$, transmits the photon to Bob, and then Bob receives the photon in certain polarization state and with an arrival time of $t^{B}_{R}$ according to his local clock $\mathbf{B}$. Similarly, Bob's transmission time is $t^{B}_{S}$, and Alice's reception time is  $t^{A}_{R}$. On the security side, there are two major classes of attacks: intercept-resend attack and time-delay attack. In intercept-resend attack, the adversary, Eve, intercepts and measures the signal and resends a new forged signal with her tailored time. To solve this attack, QSTT uses the polarization state of the single photon as the symbol to represent the time signal. Thanks to the quantum non-cloning theorem, any attempt to intercept-resend the single photon will inevitably disturb the quantum state\cite{lo2014secure}, which can be checked via the post-processing (see Step 2). The other type of attack is time-delay attack, where Eve randomly increases/decreases the propagation time of the signal during its transmission. Time-delay attack can be countered from the known space-time structure (see Step 4).

\emph{Step 2.} Alice and Bob evaluate the quantum bit error rates (QBER) in the polarization degree of freedom for the timing signals. If the QBER is below the pre-set security threshold, then QSTT succeeds; Alice and Bob proceed to generate the secret keys via QKD and post-process the timing data. Similar to QKD, the low QBER ensures the authenticity of the state of timing signals against the intercept-resend attack. Specifically, Eve will inevitably introduce a QBER of 25\% for the intercept-resend attack\cite{lo1999unconditional,shor2000simple}. Thus, for the observed QBER \emph{Q}, an upper bound of \emph{4Q} of timing signals is assumed to be tampered by Eve. In our experiment, we divide the timing signals into blocks and monitor the QBER for each block. We set a secure QBER threshold of 1.25\%, i.e., we keep only the secure blocks with $Q\leq1.25\%$. This guarantees secure data in each block with a 95\%($=1-4Q$) confidence level. Only for those secure blocks, we perform further post-processing by discarding the timing signals outside the confidence interval.

\emph{Step 3.} Alice transmits the encrypted classical timing data to Bob through a public channel, using the keys generated from the QKD. Alternatively, Bob can transmit his timing data to Alice. Note that the keys can also be generated by previous QKD rounds. QSTT uses the quantum keys to encrypt/decrypt the data to defeat the spoofing attacks on classical timing data.

\emph{Step 4.} Alice or Bob, having all the timing data, evaluates the clock offset $\tau_{BA}$ and the ranging distance $R$ with the coincidence time events of  $t^{A}_{S}$ and  $t^{B}_{R}$, $t^{B}_{S}$ and $t^{A}_{R}$ by the equations,
\begin{subequations}
\begin{align}
        \tau_{BA} &= (t^{B}_{R} + t^{B}_{S} - t^{A}_{S} - t^{A}_{R} )/2, \\
        R &= c[(t^{A}_{R} - t^{A}_{S}) - (t^{B}_{S} - t^{B}_{R} )]/2,
\end{align} \label{time-equation}
\end{subequations}
where $c$ is the speed of light. To counter the time-delay attack, either Alice or Bob needs to compare the measured distance $R$ with the prior-known distance $R_{p}$: if the distance difference $|R-R_{p}|$ is within the security alert limit $L$, then the timing data is secure and kept; otherwise, it is discarded. For this, we have made two assumptions: (a) the space-time structure near the Earth is known and cannot be changed; (b) the satellite orbit is known and cannot be modified. These assumptions ensure that the distance between the satellite and the ground station can be prior obtained securely, e.g., via satellite-orbit prediction\cite{choi2013evaluation} (see Methods). Furthermore, we use the free-space channel between satellite and ground, in which the channel is mostly in outer space and its propagation time is nearly irreducible. Notice that $c$ in Eq.~\ref{time-equation} refers to the speed of light in the space time structure near the Earth, which is assumed to be known. Together, any types of attacks to the propagation time can be countered in principle, according to Fermat's principle (i.e., principle of least time) for optical signals.

Overall, satellite-based QSTT can well satisfy the three security conditions, summarized in Table~\ref{Tab:secure conditions}. Nonetheless, the secure time transfer essentially needs the authenticity for \emph{both} the qubit state and the arrival time of the photon, which has higher security requirements than QKD\cite{lo2014secure}. Consequently, the security of QSTT is not as strong as QKD to be against any attacks. QSTT needs additional, but realistic, security assumptions to ensure the security of arrival time. Even so, our scheme already offers much higher level of security than all previous time-transfer schemes (see Supplementary Table 2). An unconditionally secure protocol and its proof may be obtained by combining the security conditions\cite{narula2018requirements} and QKD security proofs\cite{lo1999unconditional,shor2000simple}.

To verify the key technologies and show the feasibility, we performed an experimental study of satellite-based QSTT between the Micius satellite and the Nanshan ground observatory in China. In the downlink, we demonstrated the satellite-based QKD by using single photons as the carrier of time transfer, whereas in the uplink, we performed standard optical time transfer using classical laser pulses. Note that an uplink QKD, though not implemented here, is straightforward based on the technology demonstrated in satellite-based quantum teleportation\cite{ren2017ground}.

The experimental setup is illustrated in Fig.~\ref{exp_setup}. A decoy-state QKD with polarization encoding could be seen in the downlink.  Quantum signals were generated by four laser diodes at 848.6 nm with a repetition of 200 MHz and a pulse width of 200 ps. A crystal oscillator was used as the satellite clock $\mathbf{A}$, and the emission time $t^{A}_{S}$ of photon pulses was triggered by electrical signals synchronized to this clock. A 300-mm-diameter Cassegrain telescope was used to transmit these single-photon pulses to the ground, whereas a 1.2-m-diameter ground receive telescope was used to collect the QKD photons with high efficiency. The photons were coupling to a BB84 polarization-decoding module with one beam splitter and two polarization beam splitters. Finally, the photons were detected by four single-photon detectors of approximately 300-ps time jitter. The arrival time $t^{B}_{R}$ of each received photon was tagged by a time-to-digital converter.

In the uplink, a 1064 nm pulsed laser was installed on the ground station to produce signals with 0.8 ns pulse width, 15 $\mu$J pulse energy at 10-kHz repetition rate. The pulses were sampled locally and detected by a PIN diode for recording of emission time $t^{B}_{S}$ by the ground clock $\mathbf{B}$ which is an ultra-stable crystal oscillator. After transmission, the optical pulses were collected by the satellite telescope, and detected by a linear-mode-operated avalanche photodiode for recording of arrival time $t^{A}_{R}$ based on the satellite clock.

Quantum signals were transmitted and collected for about 100 s for one passage of Micius. A demonstrated satellite-based QKD could be seen in the downlink, with the time-tagged arrival time of each photon. As shown in Fig.~\ref{QKD_QBER}, we obtained a QBER of less than $1\%$. We employ decoy-state protocol and the standard error analysis to post-process the secret keys\cite{PhysRevA.72.012326} (see Supplementary Note). The final secret keys generated between the satellite and the ground amounted to 4,069,481 bits, which can be used for the encryption of the classical time-pairing data. In particular, the QKD keys were used as the seed keys for the symmetric encryption approach of advanced encryption standard (AES)-128 protocol for encryption of every 32 kB time-pairing data with 128-bit refreshed keys. The encrypted data were transmitted from the satellite to the ground via a classic microwave channel.

Accordingly, we paired the downlink and uplink time events to obtain two-way time events ($ t^{A}_{S}, t^{B}_{R}, t^{B}_{S}, t^{A}_{R}$) at a high repetition rate of nearly 9.3 kHz (see Fig. \ref{QKD_QBER}), mainly limited by the repetition rate of the uplink laser pulses. This time data rate is higher than T2L2 of 10 Hz\cite{samain2008time}. The time events were fit with the root mean square (RMS) residual. Fig. \ref{results}a and \ref{results}b show the RMS results of 310 ps and 358 ps for the downlink and uplink, respectively. Moreover, with the two-way time events, we calculated the clock offset and evaluated the distance between the ground and the satellite (see Methods). We evaluated the time transfer precision by fitting the clock offset data in every second and obtaining RMS residuals of approximately $250\sim450$ ps. The time precision $\sigma_t$ was improved to $30\sim60$ ps after we averaged 300 raw data points (see Fig. \ref{results}c), and thereby, the random error was consequently reduced. We applied the same processing method to the ranging measurement data, where the ranging’s RMS of the raw data of within $11\sim18$ cm had its precision improved to within $1\sim2$ cm after averaging.

To identify the time-delay attacks, we compared the measured ranging distance and the prior known distance, so as to guarantee that the time difference was within alert limit \emph{L}. As mentioned earlier, we experimentally achieved a ranging precision of $1\sim2$ cm with two-way optical ranging measurements. The prior known distance between the satellite and the ground can be obtained by the satellite-orbit prediction. Presently, Micius had an order of meters accuracy for orbit prediction, being limited by the performance of the satellite GPS receiver. Nevertheless, with the current technology, such accuracy can be easily improved to centimeters\cite{choi2013evaluation}. For example, the international GNSS service (IGS) provides ultra-fast products of GPS satellites of 5-cm prediction accuracy in the next 24 hours. Hence, with IGS, the alert limit can be typically set to the nanosecond or even sub-nanosecond level in the future.

Overall, we have proposed a satellite-based QSTT scheme based on two-way QKD, and experimentally verified the feasibility of QSTT in the satellite-to-ground link with sub-nanosecond precision. We expect the findings of this study to generate new possibilities toward a revolutionary quantum time-transfer network of a global scale.

\noindent \textbf{References}
\noindent
\bibliographystyle{naturemag}

\newpage

\noindent \textbf{Methods}

\noindent \textbf{Practical security for satellite-based QSTT scheme.}
\noindent Secure time transfer requires the communication security for \emph{both} physical timing signal and classical timing data. The classical timing data can be secure using the secret keys, generated from QKD, for encryption. For the physical timing signal, there are two major attacks: intercept-resend attack and time-delay attack. In intercept-resend attack, Eve violates the authenticity of the state of the signal by intercepting the signal and resends a new forged signal with her tailored time. In time-delay attack, Eve violates the authenticity of the prorogation time of the signal by randomly increasing/decreasing the propagation time of the signal during its transmission.

Satellite-based QSTT can counter both intercept-resend attack and time-delay attack by satisfying three security conditions\cite{narula2018requirements}: (i) The two stations must transmit authenticated timing signals and data; (ii) propagation time must be irreducible to within a security alert limit $L$; and (iii) the round-trip time must be known to one of the two stations to within the alert limit.

The authenticity in condition (i) means that the signal itself cannot be tampered, which aims to defeat the intercept-resend attack. QSTT uses the quantum signal, i.e., single photon, as the carrier of timing signal to ensure the unpredictability, because any attempt to intercept-resend the single photon will inevitably disturb the quantum state\cite{lo2014secure}. Condition (ii) and (iii) are primarily used to counter the time-delay attack: for a symmetry time-delay attack which increases time delay in both transmission directions, Eve can be found using condition (iii); for an asymmetry time-delay where one way increases delay and the other way reduces delay, condition (ii) is essentially needed to guarantee that the reduction in propagation time is not possible.

To satisfy condition (ii), we have assumed that the orbit of the satellite is known and cannot be modified, which can be satisfied via the satellite-orbit prediction\cite{choi2013evaluation}. In satellite-orbit prediction, the satellite acquires its own orbit information via the broadcasting of multiple GPS satellites from international GNSS service in the outer space. The tampering for GNSS service seems unlikely, since it requires the spoofing for all GPS satellites' signals in the outer space. In fact, the orbit information for most of today's GPS satellites is already public.

In QSTT, the parameter $L$ is determined by the precision of the satellite-orbit-prediction technique\cite{choi2013evaluation}, i.e., the RMS of $R_{p}$. Quantitatively, the parameter $L$ is the upper bound of the precision of the secure time transfer, i.e., $c\sigma_t\leq L$, where $c$ is the speed of the light in the known space time structure near the Earth, and $\sigma_t$ is the time precision of the secure time transfer. For each frame of data, we quantify the obtained distance $R$ via Eq.~\eqref{time-equation}, if $\mid R-R_p\mid\leq L$, then we keep the time data; otherwise we discard the data.

In satellite-based QSTT, the free-space channel mostly in outer space between satellite and ground guarantees condition (ii). Note that this condition was difficult to meet for a fiber channel, except for additional countermeasures\cite{lee2019asymmetric}. Because fiber has a larger refractive index ($\sim1.5$) than that of vacuum ($1.0$). Even so, in a rigorous manner, the optical index of the atmosphere link is not the same as that of the vacuum. But, this issue is minor in our experiment. The effective vertical thickness of Earth's atmosphere is $\sim$5-10 km, which is much smaller than Micius's altitude of $\sim$500 km. Above the Earth's atmosphere, the channel is vacuum in outer space. Moreover, the refractive index of atmosphere is $\sim$1.000273. With this value, the difference in time delay between atmosphere and vacuum over $\sim$5-10 km is $\sim$5-10 ns. In fact, such time delay is at the same level of the orbit-prediction accuracy of the Micius satellite, which can be considered in the security alert limit $L$.

\noindent \textbf{Pairing time events and calculation.} We assumed that clock $\mathbf{A}$ on the satellite was a standard clock, whereas clock $\mathbf{B}$ on the ground might have both a time and a frequency offset. The true time $t^{A}$ of clock $\mathbf{B}$ at $t^{B}$ was $t^{A}=\kappa t^{B}+\tau$, where $\kappa$ is the time scale factor of clock $\mathbf{B}$ and $\tau$ is the clock offset at $t^{B}=0$. The reference frame was chosen
so that the ground is stationary and the satellite moves at a certain speed. No
relativistic corrections were applied. The radial velocity $\nu$ of the satellite could be regarded as a constant within milliseconds when it underwent small acceleration at several tens of $m/s^{2}$. If the distance from the ground to the satellite was $R_{0}$ at time $t^{A}_{0}$, then the distance at time $t^{A}$ was $R(t)=R_{0}+\nu (t^{A}-t^{A}_{0})$. From the satellite ephemeris, we could calculate the coarse propagation delay of the signal pulses, from which we matched the emission and arrival times of a given signal pulse: $[t^{A}_{S}(i), t^{B}_{R}(i)]$ for the downlink and $[t^{B}_{S}(j), t^{A}_{R}(j)]$ for the uplink. Subsequently, we paired the downlink and uplink time events to form two-way time events $[t^{A}_{S}(i), t^{B}_{R}(i), t^{B}_{S}(j), t^{A}_{R}(j)]$, in which the emission times $t^{A}_{S}(i)$ and $t^{B}_{S}(j)$ were nearly the same. Based on the kinetic formulations, we have the following series of equations that represent the two-way time events,
\begin{subequations} \label{eq:1}
\begin{align}
    c (\kappa t^{B}_{R}(i)+\tau -t^{A}_{S}(i)) &= R_{0}+ \nu (t^{A}_{S}(i)-t^{A}_{0}) \\
    c (t^{A}_{R}(j)-\kappa t^{B}_{S}(j)-\tau) &= R_{0}+ \nu (t^{A}_{R}(j)-t^{A}_{0}).
\end{align}
\end{subequations}

Eq.~\eqref{eq:1} can be used to obtain all required parameters, including $R_{0}$, $\nu$, $\tau$, and $\kappa$. The clock offset and ranging distance from the demonstration experiment are shown in the Supplementary Fig.~1. The clock offset drift there mainly came from the frequency drift of the satellite crystal oscillator.

\noindent \textbf{Data availability}
\noindent The data represented in Figs. 3 and 4 are available as Source Data. All other data that support the plots within this paper and other findings of this study are available from the corresponding author upon reasonable request.

\noindent \textbf{Code availability}
All relevant codes or algorithms are available from the corresponding author on reasonable request.

\noindent \textbf{Acknowledgement.}
\noindent The authors thank enlightening discussions with Yuan Cao and Hoi-Kwong Lo. This work was supported by the National Key R\&D Program of China (2017YFA0303900, 2018YFB0504300), the Strategic Priority Research Program on Space Science of the Chinese Academy of Sciences, National Natural Science Foundation of China, Anhui Initiative in Quantum Information Technologies and the Key R\&D Program of Guangdong Province (2018B030325001).

\noindent \textbf{Author contributions.}
\noindent Q.Z., F.X., C.-Z.P. and J.-W.P. conceived the research. Q.S., J.Y., Y.-A. C., Q.Z., F. X., C.-Z. P. and J.-W.P. designed the experiments. F.X. analyzed the security. H.D., Q.S., C.-Z.W., S.-L.L., W.-Y.L., W.-Q.C., S.-K.L., J.-G.R., J.Y., Y.-A.C., Q.Z., F.X., C.-Z.P. and J.-W.P. developed the satellite, the payloads and the single-photon time-transfer techniques. H.D., Q.S., C.-Z.W., S.-L.L., W.-Y.L., W.-Q.C. carried out the experiment with assistance from all other authors. F.X. and J.-W.P. analysed the data and wrote the manuscript, with input from H.D., Q.S., C.-Z.W., Q.Z and C.-Z.P.. All authors contributed to the data collection, discussed the results and reviewed the manuscript.

\noindent \textbf{Competing interests.}
\noindent The authors declare no competing financial interests.

\noindent \textbf{Additional information.}
\noindent Correspondence and requests for materials should be addressed to F.X., C.-Z.P. or J.-W.P.

\clearpage

\begin{table*}
\centering
\caption{\textbf{Security conditions and satellite-based QSTT.}} \label{Tab:secure conditions}
\begin{tabular}{|p{7.5cm}|p{7.5cm}|}
      \hline \hline
      \textbf{Security conditions} & \textbf{Satellite-based QSTT} \\
      \hline
      The two stations must transmit authenticated time signal and time data. & Single photons serve as the time signals; secret keys generated via QKD encrypt/decrypt the classical time data.\\
      \hline
      The signal's propagation time must be irreducible within the alert limit. & Free-space link in outer space is adopted. \\
      \hline
      The round-trip time must be measurable to one of the two stations within the alert limit. & The link distance is prior measured via satellite-orbit prediction. \\
      \hline \hline
\end{tabular}
\end{table*}

\begin{figure}
    \centering
    \includegraphics[width=0.9\linewidth]{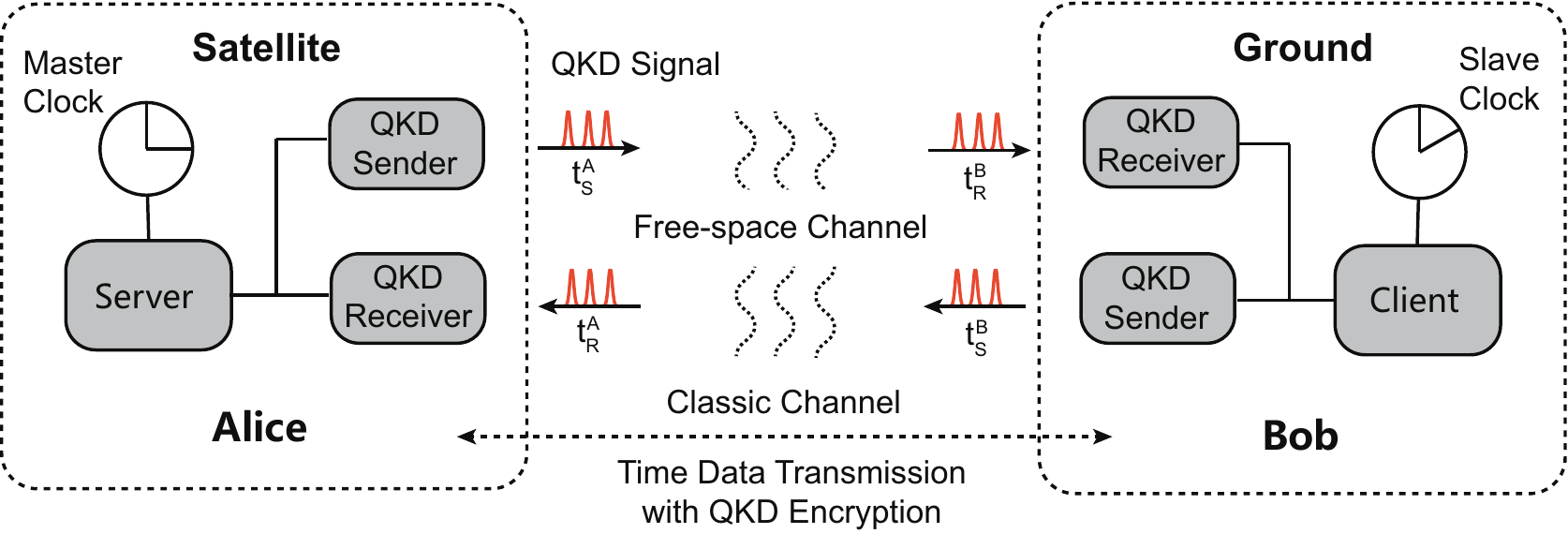}
    \caption{\textbf{Schematic diagram of satellite-based quantum-secure time transfer.} A two-way free-space QKD is established between server Alice on the satellite and client Bob on the ground. The QKD signals, encoded single photons, are used for time synchronization to prevent against malicious intercept-resend attacks to the state of the signals. Emission times $t^{A}_{S}$, $t^{B}_{S}$ and arrival times $t^{A}_{R}$, $t^{B}_{R}$ of the QKD signals are recorded with the local clock on both sides. Time data, encrypted with secret keys from QKD, are transmitted in the classical channel. Bob calculates the clock offset and link distance with these secure data, and he can synchronize with Alice's clock at high security by comparing the measured distance to the prior known distance, in order to identify the time-delay attacks.}
    \label{twoway_QKD}
\end{figure}

\begin{figure}
    \centering
    \includegraphics[width=0.9\linewidth]{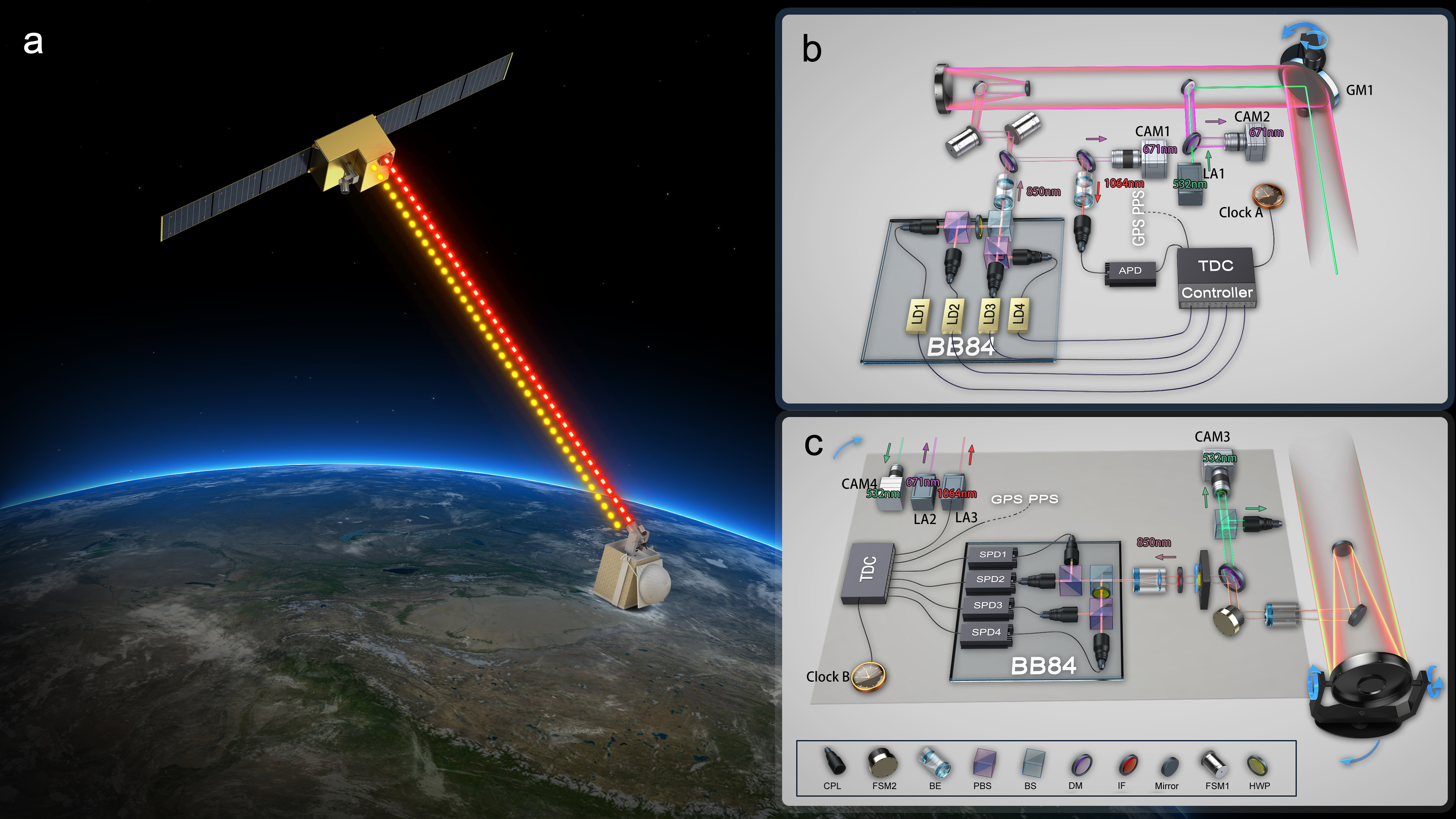}
    \caption{\textbf{The experimental setup.} \textbf{a,} An overview of satellite-based QSTT. A two-way optical link is established between Micius and Nanshan ground observatory, where the downlink uses single photons and the uplink uses classical laser pulses. \textbf{b,} Schematic of the transceiver on the satellite. The downlink single-photon signals are generated by four faint laser diodes (LD1-LD4, 850-nm), and they are encoded in BB84 polarization states. Emission time of the laser pulses is determined by clock $\mathbf{A}$ (crystal oscillator) on the satellite. The classical laser pulses from the ground (1064-nm) are detected by a linear avalanche photodiode (APD). Arrival time is tagged by the time-to-digital converter (TDC). \textbf{c,} Schematic of the transceiver on the ground. The single-photon signals are analyzed by a BB84 decoder and detected by four single-photon detectors (SPD1-SPD4). Clock $\mathbf{B}$ on the ground is an ultra-stable crystal oscillator. LA1 (532-nm laser) and LA2 (671-nm laser), and large field-of-view and fast cameras (CAM) are utilized for tracking. A GPS pulse-per-second (PPS) signal is used for initial coarse time synchronization. IF: interference filter; FSM: fast steering mirrors; CPL: coupler; BE: beam expander.
    }
    \label{exp_setup}
\end{figure}

\begin{figure}
    \centering
    \includegraphics[width=1\linewidth]{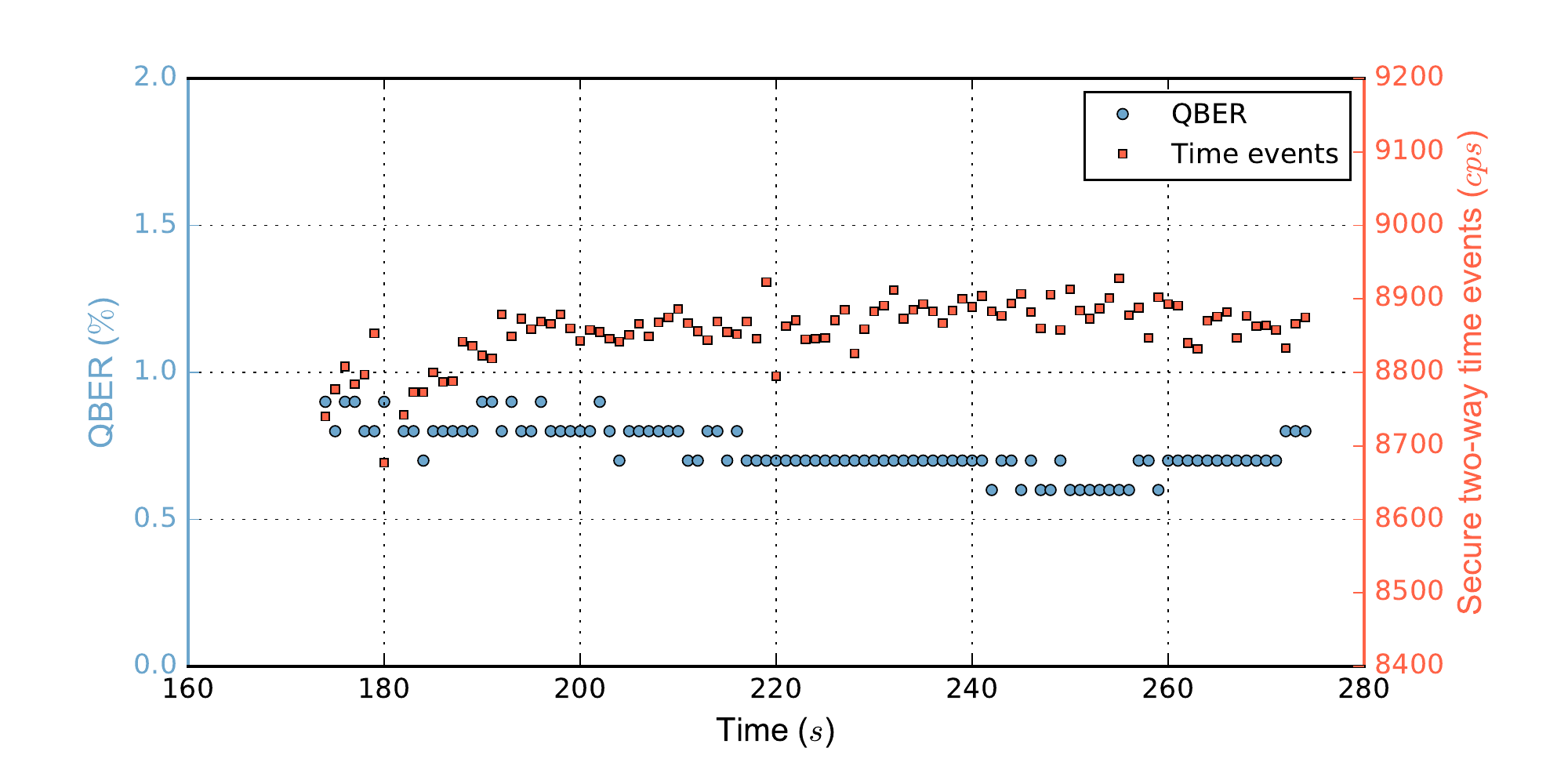}
    \caption{\textbf{QBER in the downlink and statistics of the secure two-way time events.} QBER is calculated each block of one second in the downlink QKD, presenting an average of less than $1\%$. The two-way time events are a pair of secure downlink and uplink time events which are counted by every second, and they are used for the calculations of clock offset and ranging distance.}
    \label{QKD_QBER}
\end{figure}

\begin{figure}
    \centering
    \includegraphics[width=1\linewidth]{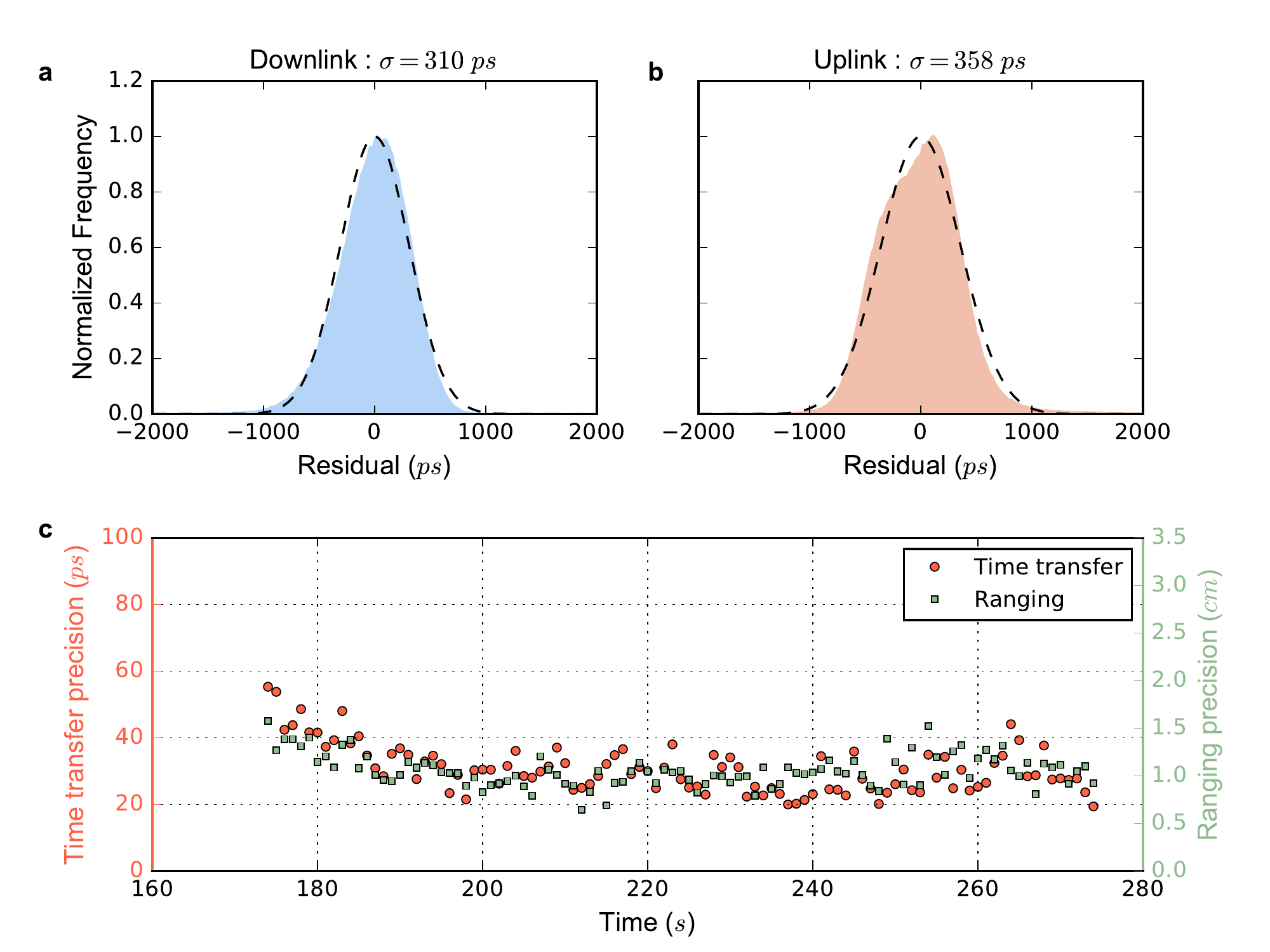}
    \caption{\textbf{Satellite-based time-transfer results.} \textbf{a,} Distribution of downlink single-photon-based time event coincidences, fit with RMS of $310$ ps. The black dashed line indicates a normal distribution. \textbf{b,} Distribution of uplink laser-pulse-based time event coincidences, fit with RMS of $358$ ps. \textbf{c,} Precision plots of time transfer and ranging. Both red and green data points are calculated from normal points that are averaged from 300 raw data results for reduced random error. On average, time transfer and ranging precision are about 30 ps and 1 cm, respectively. }
    \label{results}
\end{figure}

\end{document}